\documentclass[12pt]{article}
\usepackage{a4}
\usepackage{epsf}
\usepackage{amssymb}
\usepackage{cite}
\newcommand{\be}{\begin{equation}}
\newcommand{\ee}{\end{equation}}
\newcommand{\bea}{\begin{eqnarray}}
\newcommand{\eea}{\end{eqnarray}}

\begin{document}
\begin{titlepage}
\title{Brane Probes, Toric Geometry, and Closed String Tachyons}
\author{}
\date{
Tapobrata Sarkar
\thanks{E--mail:~ tapo@ictp.trieste.it}
\vskip0.4cm
{\sl the Abdus Salam \\
International Center for Theoretical Physics,\\
Strada Costiera, 11 -- 34014 Trieste, Italy}}
\maketitle
\abstract{We study non-supersymmetric orbifold singularities from the point 
of view of D-brane probes. We present a description of 
the decay of such singularities from considerations of the 
toric geometry of the probe branes. 
}
\end{titlepage}
\def\C{{\mathbb{C}}}
\def\R{{\mathbb{R}}}
\def\s{{\mathbb{S}}}
\def\T{{\mathbb{T}}}
\def\Z{{\mathbb{Z}}}
\def\Bbb{\mathbb}
\def\BZ{\Bbb Z} \def\BR{\Bbb R}
\def\BM{\Bbb M} \def\BR{\Bbb R}
\def\BC{\Bbb C} \def\BP{\Bbb P}
\def\CP{\BC\BP}
\section{Introduction}\label{intro}
In recent years, much attention has been paid to the study of tachyonic
instabilities in string theory. These instabilities, which usually arise
in the absence of space-time supersymmetry, have received great interest 
both in the context of open and closed strings. For open
string theories, tachyonic instabilities have been studied widely, following
the pioneering work of Sen \cite{sen}. Closed string tachyons have also received 
attention of late, following the work of Adams, Polchinski and Silverstein
(APS) \cite{aps}. Whereas open string tachyon condensation can be studied
in the boundary state formalism, and usually leads to a change in the
brane configuration (for eg. annihilation or decay of D-branes), closed string 
tachyon condensation leads to a decay of the space-time itself. This in itself 
can be a difficult problem to study, but there is a class of examples that
can be analysed with known methods. These are the non-supersymmetric orbifolds, with 
localised tachyonic instabilities, first studied in \cite{aps}. 

Resolution of supersymmetric orbifolds in closed string theory has been studied
in great details in the past. (see, for eg. \cite{asp}). In the context 
of open string theories, i.e using D-branes as probes of these orbifolds, a much richer 
structure emerges (for a review, see \cite{doug}). The papers \cite{aps},
\cite{vafa},\cite{hkmm} discusses the application of these techniques to 
non-supersymmetric orbifolds with tachyonic instabilities, with fundamentally new 
consequences (See also \cite{dghmgrp} and references therein).
 
In the typical examples studied in \cite{aps},\cite{vafa} and \cite{hkmm}, 
the action of the non-supersymmetric orbifold breaks space-time supersymmetry. 
Demanding that the (tachyonic) instabilities herein are localized at the orbifold 
fixed point, one can track the behaviour of the orbifold theory with the decay of these 
instabilities. It was found in these works, that the decay of the localised 
tachyonic instability usually drives the non-supersymmetric orbifold to a supersymmetric
configuration. In \cite{aps}, this issue was studied using D-brane probes of such 
orbifolds. Considering the world volume gauge theory of a D-brane that (lives in the 
transverse space of the orbifold and) probes the singularity, one can, using quiver 
diagram techniques developed by Douglas and Moore \cite{dm}, follow the modification of the
gauge theory as the tachyon condenses, and hence track the behaviour of 
the orbifold under tachyon condensation. 

In \cite{vafa}, this issue was considered by Vafa, who studied RG flows of 
the closed string world sheet linear sigma model \cite{wittenphases}. This was 
done for both compact and non-compact orbifold examples, 
using the mirror description of these models \cite{horivafa}. The analysis therein 
clearly validates the flow patterns discussed in \cite{aps}, and provides a 
powerful sigma-model tool to study the same. 
In \cite{hkmm}, Harvey, Kutasov, Martinec and Moore (HKMM) have 
presented a method to analyse non-supersymmeric orbifolds, using 
chiral ring techniques and the dynamics of NS 5-branes in the dual picture. 
Their method consists of the study of the chiral ring structure of the
$N=2$ superconformal field theory of the closed string world sheet. Using the 
direct correspondence of the chiral ring structure of the world sheet SCFT to 
the geometric resolution of orbifold singularities, the HKMM method is to study the 
deformation of the chiral ring along the tachyon condensation and hence derive from it 
the fate of the initial singularity along the world sheet RG flow effected by
the tachyon condensation. For non-supersymmetric orbifolds, HKMM defined the quantity 
$g_{cl}$, the coefficient in the expression of the asymptotic density of states 
in the CFT, and conjectured that this decreases along the RG flow, while 
leaving the effective central charge of the theory unchanged. 

It is of interest to continue these investigations along the lines of
APS, Vafa and HKMM, to understand the more general underlying structure of these
non-supersymmetric orbifolds, and their fate under closed string tachyon 
condensation. We expect tools from toric geometry, associated with the D-brane probe 
theory to be useful in this study. Namely, using toric geometry techniques, one can 
hope to understand the behaviour of higher dimensional non-supersymmetric orbifolds 
(for which a canonical resolution is not available) under tachyonic decay. A related 
issue that one might address is whether this decay of space-time is more generic. 
Namely, given a generic background which breaks space-time supersymmetry, is there a 
process in string theory itself, which would lead to a decay of this background into a 
supersymmetric string background. 

It is these issues that we address in this paper. As a first step, we partially 
generalise the D-brane probe results of APS for non-supersymmetric 
orbifolds of the form $\BC^2/\Gamma$. From the probe point of view, the 
D-brane can see only certain types of decays, and we address the question of 
the classification of such theories that flow in the IR to supersymmetric orbifolds. 
Next, we study the chiral ring techniques of HKMM as applied to non-supersymmetric 
orbifolds of the form $\BC^2/\Gamma$, which are related to the issue of these decays 
seen from the point of view of toric geometry in two complex dimensions. 
Our interest is two-fold. Firstly, a toric geometry picture of the decay process 
for two-fold orbifolds would be an useful tool in understanding such processes in 
higher dimensions. Secondly, one would, via this picture, be able to study, for 
example, the decay of generic weighted projective spaces using the picturisation of 
D-branes in weighted projective spaces as D-branes on the resolutions of 
(higher-dimensional) orbifolds \cite{diadoug}. This study has already been initiated in 
\cite{vafa}, and our methods are complimentary to the mirror symmetry principles used there.
Further, we study the inverse toric procedure pioneered in \cite{fhh}
for some simple examples. This procedure is expected to play an important role in the 
full understanding of the decay of non-supersymmetric backgrounds, as we will point out 
in the paper. 

This paper is organised as follows. In section 2, we briefly review the probe brane 
analysis of APS, and study some flow patterns for the $\BC^2$ orbifold. 
Section 3 begins with a brief review of the chiral ring analysis of HKMM, after which
we discuss the toric geometry of the D-branes probing non-supersymmetric orbifolds, 
and discuss several examples, both for two and three-fold orbifolds. Next, we study the 
inverse toric procedure, applied to non-supersymmetric orbifolds and initiate a discussion 
on toric duality for non-supersymmetric orbifolds. Section 4 ends with some 
discussions and conclusions. 
 
\section{D-brane Probes of Non-Supersymmetric Orbifolds}

In \cite{aps}, APS has studied closed string theory on non-supersymmetric orbifold
backgrounds. As we have mentioned before, these break space-time supersymmetry, 
and have tachyonic modes in (some of) the twisted sectors. An important aspect of 
the theories that have been studied is that the tachyonic excitations are localised
at the fixed points of the orbifolds, and do not affect the stability of the
bulk space-time. It was shown in \cite{aps} that these orbifolds decay with time, 
and the final theory reached via this decay process is a supersymmetric orbifold. 
There are two distinct scales involved in this problem. First, one can study the decay 
process in the sub-stringy regime, where the tachyon expectation value is small. 
Here, one expects the world volume gauge theory of D-branes probing these orbifolds 
to provide an useful tool in studying the decay process. In the substringy regime, 
one can study the world volume gauge theory of a D-brane that probes the 
non-superymmetric orbifold, using quiver diagram techniques
developed in \cite{dm}. (Here, one is dealing with the classical worlvolume
gauge theory that lives on the branes). Far from the substringy regime, 
when $\alpha'$ corrections become large, the probe analysis is not suitable
any more, and one has to revert to a gravity analysis. 

These two approaches were studied in \cite{aps} for the case of orbifolds of
the form $\BC/\BZ_n$. It was shown that these non-supersymmetric orbifolds 
decay to orbifolds of lower rank, and the process continues until one reaches
flat space. Similarly, for $\BC^2/\Gamma$ orbifolds, brane probes lead to the 
prediction of transitions from non-supersymmetric orbifolds to supersymmetric 
orbifolds (of lower rank). The probe analysis is again done for the sub-stringy regime. 
The method of \cite{aps} is to excite marginal deformations in the original theory, 
which takes the system to a lower rank orbifold, which is only locally supersymmetric. 
Deformations of this latter theory, which are expected to be tachyonic
in nature, then drives the system to a supersymmetric configuration. 

It is important to turn on only marginal perturbations in this method of
probing a non-supersymmetric orbifold. If one turns on generic 
tachyonic deformations, quantum corrections will become important, and
the the classical brane probe theory ceases to be useful. In exciting
marginal deformations in the theory, one has to maintain a certain
quantum symmetry out of the full symmetry group. This quantum symmetry is 
retained by the D-brane theory once one breaks the other part of the 
orbifold group. We will now elaborate on this in some more details. 

\subsection{General Pattern for Quivers}
 
Let us begin by considering D-brane probes of Type II string theory on 
the $\BC^2$ orbifold based on the general twist
\begin{equation}
R = {\mbox{exp}}\{\frac{2\pi i}{n}\left(J_{67}+kJ_{89}\right)\}
\label{p1}
\end{equation}
where $J_{67}$ and $J_{89}$ refer to the rotations in the complex planes
$Z^1=X^6+iX^7$ and $Z^2=X^8+iX^9$. 
We can consider a D-p brane probe of the above geometry, where the brane extends only 
along the transverse directions. The low energy theory of such a configuration is the 
orbifold of the $N=4$ world volume gauge theory on the D-brane, with the usual 
\cite{dm} projection conditions.  Following the notation of \cite{aps}, we call this 
orbifold $\BC^2/\BZ_{n(k)}$.  The world volume spinors in the ${\bf 16}$ of $SO(9,1)$, 
$\eta$ and $\chi$, are labelled by their weights under $SO(4)$, and the $SO(5,1)$ 
spinor indices are suppressed \cite{aps}. In this notation, $\eta$ is the $(-,+)$ 
component, and $\chi$ is the $(-,-)$ component.
 
The quiver diagram is obtained by following the by now standard prescription due
to Douglas and Moore \cite{dm}, and will have $n$ nodes, corresponding to the 
$n$ $U(1)$ factors of the gauge group $U(1)^n$. As we have mentioned, an orbifold of this 
form will not preserve space-time supersymmetry, and will flow (in the sense of 
the RG) to a supersymmetric orbifold via the condensation of twisted sector
tachyons. It is an interesting question to classify these flows using general 
quiver techniques, and we will present some results on the decay of non-supersymmetric
orbifolds by turning on marginal deformations. A complete classification of generic
flows using the methods of \cite{aps} is a difficult issue to address. For the purpose
of this paper, we will make the simplifying assumption of turning on only marginal 
perturbations, but, as we will see later in the paper by using tools from toric 
geometry, there are several interesting aspects of such flows.   

Let us begin by reviewing the procedure due to APS for the decay of a non-supersymmetric 
orbifold singularity. As we have already mentioned, the essential idea 
is to turn on marginal or tachyonic deformations from a given twisted sector, which 
is expected to produce a partial blowup of the initial singularity. Once the system 
has reached such a stage, one can consider turning on further deformations which are 
tachyonic in nature, and drives the system to a supersymmetric configuration. 

In terms of the analogues of the F and D terms that appear in the D-brane world 
volume gauge theory in the supersymmetric case, this is tantamount to turning on 
certain Fayet-Illiapoulos (FI) parameters, in a way that a quantum symmetry is maintained 
at the end.  After choosing a particular vaccum, in which we gives vev's to a certain 
set of fields maintaining this symmetry, we are left with a reduced world volume theory 
(integrating out fields that become massive due to the vev's) that, upon suitable 
rearrangement, can be seen to correspond to a different (lower rank) orbifold action.

In particular, from eq.(\ref{p1}), one can reach a configuration
that preserves a $\BZ_m$ symmetry, where $m$ divides $n$. This might be
achieved by turning on marginal deformations from an appropriate twisted
sector. 

There are two distinct choices of vaccum corresponding
to turning on vevs for the fields that parametrise either of the two 
$\BC^2$ directions. In either case, the final symmetry group that is restored
will depend on the symmetry of terms that take vev's. It turns out that the
analysis of the vacua that preserves an $m$ fold symmetry coming from the 
vevs of the $Z^1$ is qualitatively different from the corresponding vacua
where one chooses the $Z^2$ vevs. Let us now study this in some details. 

\subsection{Reaching a supersymmetric configuration}

In what follows, we will study a class of non-supersymmetric
orbifolds of the form $\BC^2/\BZ_{n(k)}$ that flow to orbifolds of the 
form $\BC^2/\BZ_{m(k')}$ where $m$ is a factor of $n$. Flows starting from the
non-supersymmetric orbifold of the form $\BC^2/\BZ_{2l(2l-1)}$ and $\BC^2/\BZ_{2l(3)}$ 
have been considered in \cite{aps} (with $m$ being $2$ and $l$ respectively). In general, 
the final supersymmetric configuration will be of the form $\BC^2/\BZ_{m(1)}$ 
\cite{aps}, and will have the interpretation of having an opposite supersymmetry 
from the usual $\BC^2/\BZ_m$ orbifold. 
How does a D-brane see such a decay ? Let us take an example. Consider the
orbifold $\BC^2/\BZ_{8(3)}$. The quotienting group $\BZ_8$ has the subgroups
$\BZ_2$ and $\BZ_4$. Taking the second or the fourth power of $R$ in eq. (\ref{p1}), 
we obtain
\begin{equation}
R^{2(4)}~=~{\mbox{exp}}\{\frac{2\pi i}{4(2)}\left(J_{67}+3J_{89}\right)\}
\end{equation}
Taking the fermionic part of the string theory into account, $\BC^2/\BZ_{2(3)}$ 
(or equivalently $\BC^2/\BZ_{2(-1)}$ is a supersymmetric background, whereas 
$\BC^2/\BZ_{4(3)}$ is not. Hence, only a deformation by $R^4$ will be marginal. 
If we turn on the
marginal deformation corresponding to the fourth subsector of the theory, 
we get, as the end product, the orbifold $\BC^2/\BZ_{4(1)}$ \cite{aps}. 

The procedure of APS is to generate vevs for the fields of the theory
in such a way as to maintain a certain subgroup of the initial orbifold group, 
corresponding to the turning on of an appropriate twisted sector. The 
vev breaks the other part of the group action, and we are left finally with the subgroup 
that we had maintained in choosing the vevs. Of course, one might expect that such a 
constraint is not necessary. Namely, by choosing arbitrary vevs for certain fields 
in the D-brane gauge theory, one might still reach a supersymmetric 
configuration. Such an example, analysed in \cite{aps} is the decay of the orbifold 
$\BC^2/\BZ_{5(2)}$ to a supersymmetric configuration. Here, the deformations are 
entirely tachyonic, there being no marginal deformations in any of the twisted sectors. 
As we have pointed out, it is difficult to classify completely such generic deformations
using the methods of this subsection, and we will not treat this issue here.

Let us concentrate on the cases where the quotienting
group admits of discrete subgroups, i.e, the cases $\BC^2/\BZ_{n(k)}$ where
$n$ admits of factors greater than unity. By turning on marginal perturbations that
corresponds to maintaining a discrete subgroup of the initial quotienting group, 
we can reach supersymmetric configurations. This restricted class of flows can be
classified by noting that since the final configuration is guaranteed to be 
of the form $\BC^2/\BZ_{n'(1)}$, we expect loops in the final (annealed) quiver diagram, 
arising out of the spinors $\eta$ or ${\bar \eta}$, as can be seen by inspecting their
spin. Therefore, we need to choose the vaccum of the original theory in such a way that 
these loops are produced in the final quiver diagram. In general, if there are several 
supersymmetric sectors in a non-supersymmetric orbifold, the final supersymmetric 
configuration might be reached in several steps, we will return to this question in a while. 

\subsection{Turning on VEVs for the $Z^1$ and $Z^2$}

For the orbifold action of eq. (\ref{p1}), the surviving components of the 
coordinates $Z^1~(X^6+iX^7)$ after projection by the orbifolding group takes the 
form $Z^1_{ij}$ where 
\begin{equation}
(i,j):~~(1,2),~(2,3),~\cdots(n-1,n)
\end{equation}
Now, we wish to turn on marginal deformations so that the symmetry group $\BZ_m$ 
is maintained. This would involve identifying the fields under an $m$ fold symmetry 
and choosing a vaccum that restores this symmetry. For example, one set of the fields 
$Z^1_{ij}$ that are identified, are, 
\begin{equation}
(ij):(1,2),\left(1+\frac{n}{m},2+\frac{n}{m}\right),
\cdots\left(1+(m-1)\frac{n}{m},2+(m-1)\frac{n}{m}\right)
\label{vevsz1}
\end{equation} 
and similar identifications hold for other sets of fields. 
In order to maintain an $m$ fold symmetry in the final orbifold theory,
one choice of the massless fields is given by $Z^1_{\frac{n}{m}j,\frac{n}{m}j+1}$,
with the rest of the $Z^1$ components acquiring vevs. Now, in order to determine the 
massless fermions, we need to inspect the Yukawa terms, which are of the form 
$L_Y={\mbox{Tr}}\{\left[Z^1,\chi\right]\eta+\left[Z^2,\chi\right] {\bar \eta}+h.c\}$
\cite{dm},\cite{aps}. It suffices to consider the first term in this case, and with
our choice of the massless components of $Z^1$, the massless components of $\eta$ 
are determined from the term 
\begin{eqnarray}
\left[Z^1_{\frac{n}{m}j,\frac{n}{m}j+1}\chi_{\frac{n}{m}j+1,\frac{n}{m}j
+\frac{2n-k+1}{2}}
-\chi_{\frac{n}{m}j,\frac{n}{m}j
+\frac{2n-k-1}{2}}Z^1_{\frac{n}{m}j+\frac{2n-k-1}{2},\frac{n}{m}j+\frac{2n-k+1}
{2}}
\right]\nonumber\\ 
\times\left(\eta_{\frac{n}{m}j+\frac{2n-k+1}{2},\frac{n}{m}j}\right)
\label{etaloop}
\end{eqnarray}
Here $j$ is an integer, which, without loss of generality, we can choose
to be unity. First, notice that in order to get a sensible final quiver where the 
massless $\eta$'s arise as loops, we require the remaining fermionic fields (of the final 
fermion quiver) to arise as massless linear combinations of the $\chi$'s. (In general,
such massless combinations will appear with the massive components of $Z^1$). 
For these to occur, both the components of $Z^1$ appearing in (\ref{etaloop}) need to
be massless, and we see from index matching that such massless combinations will 
occur only when
\begin{equation}
\left(k+1\right)~=~2p\frac{n}{m}
\label{p8}
\end{equation}
where $p\in \BZ$. This condition is actually the same as that for the existance of 
supersymmetric subsectors in the sector twisted by $m$, as can be seen by taking
the $m$th power of $R$ in eq. (\ref{p1}). 
To obtain loops for the $\eta$'s in the final quiver, we see by inspecting the 
indices of the $\eta$ fields in eq. (\ref{etaloop}), that the following constraint has to 
be satisfied 
\begin{equation}
\frac{n}{m}+\frac{1}{2}\left(1-k\right)=b\left({\mbox{mod}}~n\right)
\label{r1}
\end{equation}
where $b=1,2,\cdots\frac{n}{m}$. (We can see this by setting $j=1$ in (\ref{etaloop}),
and noting that the bosonic vevs identify the nodes $1,2,\cdots\frac{n}{m}$ in the original
quiver diagram). Combining (\ref{r1}) and (\ref{p8}), we obtain the condition
\begin{equation}
\left(1-p\right)\frac{n}{m}+1=b\left({\mbox{mod}} n\right)
\end{equation}
Which is satisfied by $p=1$, which, as we will see, will be the case for most 
of our examples. 

From the above discussion, it follows that the following flow is possible by turning 
on the $Z^1$ vevs
\begin{equation}
\BC^2/\BZ_{jl(2j-1)}\rightarrow\BC^2/\BZ_{l(1)}
\label{p10}
\end{equation}
here, $p=1$ from eq. (\ref{p8}), and eq. (\ref{r1}) is satisfied.   
As a special case of this equation, we see the flow patterns
from the above equation
\begin{eqnarray}
\BC^2/\BZ_{2l(3)}\rightarrow\BC^2/\BZ_{l(1)} \nonumber\\
\BC^2/\BZ_{2l(2l-1)}\rightarrow\BC^2/\BZ_{2(1)}
\end{eqnarray}
which have been considered in \cite{aps}.

Let us now consider the example of a non-supersymmetric orbifold that has more than 
one supersymmetric subsector. Consider, for example,
the flow 
\begin{equation}
\BC^2/\BZ_{12(7)}\rightarrow\BC^2/\BZ_{3(1)}
\end{equation}
This satisfies (\ref{p8}) with $p=1$, and corresponds to turning on marginal 
deformations from the third twisted sector. However, one could consider turning on 
marignal deformations from the sixth twisted sector, which is also supersymmetric. 
Marginal deformations from this sector, however, does not make the final orbifold 
supersymmetric, because, for a final $\BZ_6$ symmetry, we find that equation (\ref{r1}) 
is not satisfied with $n=12$ and $m=6$ ($p=2$ in this example). It can be checked 
that this flow is 
\begin{equation}
\BC^2/\BZ_{12(7)}\rightarrow\BC^2/\BZ_{6(3)}
\end{equation}
The orbifold on the r.h.s will further decay into the supersymmetric
orbifold, $\BC^2/\BZ_{3(1)}$. This is an example of a double-decay
process. Some other examples of such decays, by turning on vevs for the
$Z^1$ are 
\begin{eqnarray}
\BC^2/\BZ_{12(11)}\rightarrow\BC^2/\BZ_{4(3)}\rightarrow\BC^2/\BZ_{2(1)}
\nonumber\\
\BC^2/\BZ_{12(11)}\rightarrow\BC^2/\BZ_{6(5)}\rightarrow\BC^2/\BZ_{2(1)} 
\end{eqnarray}
Here, from the initial orbifold, by exciting the fourth and the sixth twisted sectors 
(both of which have marginal deformations), one can reach an identical configuration, 
via two different routes. The last stage of the decay, in both the cases, satisfies 
the condition in eq. (\ref{r1}) with $p=1$. 

To summarize the discussion so far, we have seen that for orbifolds which
have multiple supersymmetric twisted sectors (in the sense of \cite{aps}),
a marginal deformation that drives the orbifold into a supersymmetric one
must obey the condition in eq. (\ref{r1}). If we excite 
a marginal deformation that does not satisfy these equations, the orbifold flows
into a non supersymmetric orbifold of lower rank, from which it finally decays, by
marginal deformations, into a supersymmetric orbifold, and in the final step, 
the conditions in eq. (\ref{r1}) is satisfied.
 
\begin{figure}
\centering
\epsfxsize=3.5in
\hspace*{0in}\vspace*{.2in}
\epsffile{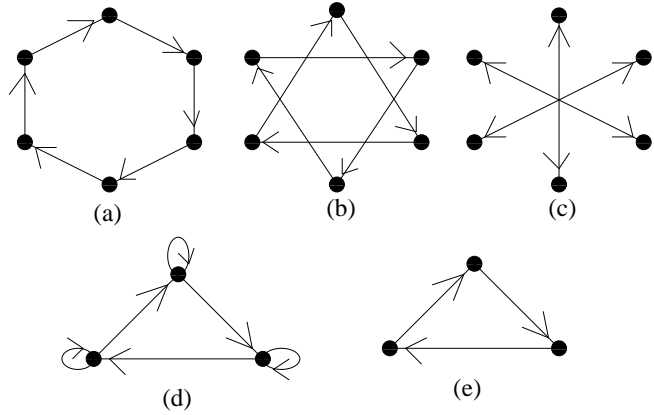}
\caption{\small Quiver Diagram for the process $\BC^2/\BZ_{6(7)}\rightarrow
\BC/\BZ_{3(1)}\times C$. (a) shows the initial bosonic quivers for the two
worldvolume scalars, (b) and (c) are the quivers for the $SO(9,1)$ fermions.
(d) shows the final quiver for the bosons and (e) shows the final quiver for 
both the fermionic fields.} 
\label{fig1}
\end{figure}

Let us now consider another class of examples where an $l$ fold final symmetry is 
preserved starting from a $2l$ fold symmetry, but for which eq. (\ref{p10}) is not 
satisfied.   For eg. consider the orbifold $\BC^2/\BZ_{2l(2l+1)}$ for odd $l$.  
In this case, after effecting the orbifold projection in the D-brane gauge theory, 
the surviving components of the $Z^1$ and $Z^2$ are both of the form $Z^i_{j,j+1}$
for $j=1,2,\cdots 2l$. The components of the $SO(9,1)$ spinors which survive, 
are $\chi^{(--)}_{j,j+l-1}$ and $\eta^{(-+)}_{j,j+l}$. If we now give vevs to 
$Z^1_{2j-1,2j}$, the bosonic fields that remain massless, are $Z^1_{2j,2j+1}$ and 
$Z^2_{2j-1,2j}$. Similarly, the fermion components that remain massless can be shown to be
$\eta_{2j-1,2j-1+l}$ and the linear combinations of the $\chi$ fields, namely
$\left(\chi_{2j-1,2j+l-2}+\chi_{j2,2j+l-1}\right)$. With these, the final orbifold is
seen to be non-supersymmetric, and of the form form $\BC/\BZ_l\times \BC$. This will 
finally decay into flat space via further twisted sector tachyon condensation as discussed 
in \cite{aps}. In summary, the flow pattern just discussed is
\begin{equation}
\BC^2/\BZ_{2l(2l+1)}\rightarrow\BC/\BZ_l\times \BC
\end{equation}
The initial and final quiver diagrams for one such process, 
$\BC^2/\BZ_{6(7)}\rightarrow\BC/\BZ_{3(1)}\times C$ is shown in figure (\ref{fig1}).

A similar analysis can be performed to study the flow of non-supersymmetric
orbifolds upon turning on the vevs of the $Z^2$ fields. As before, we can study
the index structure of the surviving fields in order to determine the flow patterns. 
As has already been noted in \cite{aps}, turning on the $Z^2$ vevs will, in general, 
drive the initial orbifold to a configuration different from the one reached 
with corresponding $Z^1$ vevs. For example, the non-supersymmetric orbifold 
$\BC^2/\BZ_{2l(3)}$ has the following flow pattern
\begin{equation}
\BC^2/\BZ_{2l(3)}\rightarrow\BC^2/\BZ_{l(1)}\oplus\BC^2/\BZ_{l(-3)}
\label{contra}
\end{equation}
The two terms on the r.h.s of the above equation are obtained by marginal deformations 
from the $l$th twisted sector by turning on vevs for $Z^1$ and $Z^2$ respectively. 
Whereas the flow effected by the $Z^1$ vev has, as the endpoint, a supersymmetric orbifold, 
that by the $Z^2$ vev is non-supersymmetric for $l>2$ (for a possibility of such 
transitions for higher values of $l$, see \cite{hkmm}). In general, for non-supersymmetric 
orbifolds of the form $\BC^2/\BZ_{n(l)}$, the analysis of the index structure of the 
surviving fields have to be done case by case. The results, we believe, are not very 
illuminating, considering the fact that such analysis as presented in this subsection can 
be done with relative ease only for cases where marginal deformations are turned on, thus 
restricting its applicability. We will therefore proceed to study the D-brane gauge 
theory on non-supersymmetric orbifolds, using chiral ring methods and toric geometry. 
However, before ending this section, let us point out that a similar quiver diagram 
analysis can be done for D-branes probing orbifolds of the form $\BC^3/\BZ_{n(k,m)}$ 
where $k$ and $m$ are integers that arise in the orbifold action
\begin{equation}
R={\mbox{exp}}\frac{2\pi i}{n}\left(J_{45}+kJ_{67}+mJ_{89}\right)
\label{threefold}
\end{equation}
As in the two-fold example, constrains on $k$ and $m$ arise due to conditions
of localisation of the tachyon. We will briefly mention this class of examples
in the next section.

\section{Chiral Ring Techniques and Toric Geometry Methods}

In the NS-R formalism, the string world sheet conformal field theory has ${\cal N}=2$ 
supersymmetry, and is endowed with ring of (anti)chiral primary operators in the NS 
sector. By extending the 1-1 correspondence between the chiral ring and the geometric 
blow-up modes to relevant perturbations \cite{hkmm}, the decay of non-supersymmetric 
orbifolds can be studied from a geometric point of view.

To summarize the construction of the chiral ring \cite{dixon}\cite{hkmm}, we recall 
that for theories on orbifolds of the form $\C^d/\Z_n$ there is one twisted sector 
associated with each element $g^j \in \Z_n$ , $j = 1,2,..,n-1$ , $g^n=1$. In 
each twisted sector (for each complex dimension which is orbifolded), there is a  
bosonic and a fermionic twist operator that can be combined to form the building 
blocks for the chiral operators of the worldsheet theory.  Bosonizing the fermionic 
fields as $\psi_i=e^{H_i}$, the twisted sector chiral operators can be written in the 
case of two-fold orbifolds of the form $\BC^2/\BZ_{n(k)}$ as \cite{hkmm}
\begin{equation}
X_j=X_j^{(1)}X_{n\{\frac{jk}{n}\}}^{(2)},~~~~~
X_j^{(1)}=\sigma_{\frac{j}{n}}{\mbox{exp}}\left[i\left(\frac{j}{n}\right)
\left(H_1-{\bar H_1}\right)
\right] 
\end{equation}
where $(1)$ and $(2)$ denote the two complex directions that are orbifoldized, 
$\sigma_{\frac{j}{n}}$ is the bosonic twist $j$ operator \cite{dixon},
and in the above equation, a similar expression holds for the operator $X^{(2)}$ as
for $X^{(1)}$. As usual, $j=1,\cdots n-1$ label the twisted sectors, and $\{x\}$ 
denotes the fractional part of $x$. The R-charges (with respect to the world sheet 
$U(1)$ current) of these operators are given by 
\begin{equation}
R_j = \frac{j}{n} + \{\frac{jk}{n}\}
\end{equation}
Now, the chiral GSO projection for Type II strings acts on the bosonised fermions as
\begin{equation}
H_1 \rightarrow H_1 - k\pi~; ~~~H_2 \rightarrow H_2 + \pi
\end{equation}
Considering the fact that in the untwisted sector of the orbifold, this must
reduce to the standard $(-1)^{F_L}$, $k$ is fixed to be odd as in the D-brane probe 
analysis of \cite{aps}. Marginal deformations of the CFT correspond to 
perturbations by operators which have $R_j=1$ \cite{dixonictp}, and 
there is an 1-1 correspondence of these with the blow-up modes of the orbifold. This 
correspondence can be extended \cite{hkmm} to the relevant modes ($R_j<1$) which
correspond to tachyonic excitations in space-time, by using a toric geometry description 
for the singularity. However the R-charge being non-integral, the spectral flow 
argument for the correspondence does not hold here. 

Several flow patterns have been analysed in \cite{hkmm}, and the $g_{cl}$ conjecture has 
been verified for these. In the brane probe analysis of the last section, we restricted 
ourselves to considering only marginal deformations of the D-brane theory. Whenever this
arises as special cases in the analysis of HKMM, the results are seen to agree. 

The above formalism can be generalised to the case of 3-fold orbifolds, where
the action of discrete group is as in (\ref{threefold}). For the case of 3-fold
orbifolds, however, a canonical resolution does not exist, unlike the two-fold 
examples. We will consider the 
simplest class of examples for these three-folds in a while, in which we 
set, in eq. (\ref{threefold}), $k=1$. These non-supersymmetric three-fold orbifolds
will, in general, have tachyonic excitations, and, following \cite{aps} or 
\cite{hkmm}, the condition for the localisation of tachyons in these cases 
can be shown to imply that the integer $m$ in (\ref{threefold}) is even (with
$k=1$). As 
in \cite{hkmm}, one can again construct the chiral ring and analyse the structure 
of flows. We will, however, perform an equivalent analysis for these cases in terms 
of the toric geometry of the probe branes in the next subsection. 

\subsection{Using Tools From Toric Geometry}

We now explore the above processes exemplifying the decay of space-time by 
using methods of toric geometry. As we have already mentioned, toric geometry is 
an extremely useful tool in studying singular spaces. However, as a caveat, we note 
that the method, which we will now elaborate, concerns only the bosonic subsector 
of the string theory. This is not a problem however, since the corresponding fermions 
can be introduced at any stage of the calculation. The methods that we will
use are fairly standard in the mathematics literature 
\cite{fulton},\cite{oda}. For the relevant physics, the reader is referred
to \cite{agm},\cite{greenerev} and references therein. 

We will deal with orbifolds of the form $\BC^m/\Gamma$, where $\Gamma$
is the discrete group $\BZ_n$. For the moment, we concentrate on the case
$m=2$, where the action on the complex coordinates 
of the discrete group on $\BC^2$ is given by
\begin{equation}
\left(Z_1,Z_2\right) \rightarrow \left(\omega Z_1,\omega^k Z_2\right)
\end{equation}
where $\omega~=~{\mbox{exp}}[2\pi i/n]$ is the $n$ th root of unity, and 
$k$ is an integer, with $|k| < n$. This is the Hirzebruch-Jung
singularity, and following the notation of APS, this
singularity is denoted by $\BC^2/\BZ_{n(k)}$.

The toric variety for the correspoding non-singular 
(i.e resolved) geometry is given by specifying a set of lattice points in 
the two dimensional lattice (with $SL(2,\BZ)$ automorphism) generated by the 
unit vectors $(1,0)$ and $(0,1)$, which we call $e_1$ and $e_2$.  
Given a singularity of the form $\BC^2/\BZ_{n(k)}$, the 
toric diagram of its minimal resolution is given by the cone generated by two vectors,
$v_f~=~e_2$ and $v_i~=~\left(ne_1 - ke_2\right)$. It can be shown \cite{fulton},\cite{oda} 
that there are $r$ added vertices inside the cone (corresponding to the
blow up modes) generated by $v_f$ and $v_i$, and these are determined from the relations
\begin{equation}
a_iv_i~=~v_{i-1}~+~v_{i+1}
\end{equation}
where the coefficients $a_i \geq 2$ and are the integers appearing in the
Hirzebruch-Jung continued fraction
\begin{equation}
{n\over k}=a_1 - \frac{1}{a_2-{\frac{1}{\cdots~- \frac{1}{a_r}}}} 
\end{equation}
where it is understood that for $k\equiv n+k$ for $k < 0$. 
Each interior vector is an exceptional divisor, and correspond to the blowing
up of $\BP^1$s, with self intersection number $-a_i$. We will follow the
standard notation, where the continued fraction is denoted by
$[a_1,a_2,\cdots~,a_r]$.

Let us take the example of the supersymmetric orbifold $\BC^2/\BZ_{4(-1)}$, 
considered in \cite{hkmm}. The corresponding (non-singular) toric variety is generated 
by the set of five vectors, which are, $(0,1),(1,0),(2,-1),(3,-2),(4,-3)$. 
Note that the toric data of this orbifold is the same as that for $\BC^2/\BZ_{4(3)}$. 
The latter is not space-time supersymmetric. We will keep this in mind, it being
implied that whether the orbifold is supersymmetric or not can be checked
by taking into account the fermionic quiver diagram. The toric data can be
arranged in an array
{\small
\begin{eqnarray}
{\cal T}=
\pmatrix{
1&0&-1&-2&-3\cr
0&1&2&3&4\cr
}\label{data1}\end{eqnarray}}

The continued fraction for this example is $[2,2,2]$. 
In \cite{hkmm}, it was shown that perturbing the Lagrangian of the closed string
theory probing this orbifold by a chiral primary operator, 
and taking the coupling of this operator
to be very large, corresponds to a splitting up of the space. Noting that
the perturbation corresponds to the blowing up of an appropriate 
$\BC\BP^1$, and the operation of taking the coupling to infinity is to
effectively blow up this $\BC\BP^1$ to infinite size (and hence to decouple
it from the geometry), the splitting is denoted by
\begin{equation}
\BC^2/\BZ_{n(-1)}\rightarrow \BC^2/\BZ_{j(-1)}\oplus \BC^2/\BZ_{n-j(-1)}
\label{split1}
\end{equation}

It is easy to understand this process from toric diagrams. Splitting up the 
toric cone will correspond to a split of the toric data into 
two parts, using any one interior vector twice. For example, the data
in (\ref{data1}) can be split into 
{\small
\begin{eqnarray}
\pmatrix{
1&0&-1\cr
0&1&2\cr}~~~{\mbox{and}}~~~
\pmatrix{
-1&-2&-3\cr
2&3&4\cr}
\end{eqnarray}} 
The first matrix can be recognised to be the toric data of the resolution
of the orbifold $\BC^2/\BZ_{2(1)}$. Using the automorphism of the two dimensional
lattice, the second matrix, after a transformation
by the $SL(2,Z)$ matrix $\pmatrix{3 & 2 \cr -2 & -1}$, can also be brought into 
the form $\pmatrix{1&0&-1\cr 0&1&2}$. This is the analogue of the 
flow pattern of (\ref{split1}), namely, 
\begin{equation}
\BC^2/\BZ_{4(-1)}\rightarrow\BC^2/\BZ_{2(-1)}\oplus\BC^2/\BZ_{2(-1)}
\end{equation}

Let us point out at this stage that deforming the CFT by marginal operators will,
in general, correspond to splitting the toric data along a vector that lies on the edge
of the toric cone connecting $v_i$ and $v_f$. We will come back to this later. 
Let us now take another example. Consider, for eg., the orbifold $\BC^2/\BZ_{10(-3)}$, 
which is non-supersymmetric. The toric data for this orbifold is given by
\begin{equation}
{\cal T}~=~\pmatrix{1&0&-1&-2&-7& \cr 0&1&2&3&10}
\label{data2}
\end{equation}
corresponding to the continued fraction $[2,2,4]$. As shown in \cite{hkmm},
deforming by the generators of the closed string CFT corresponding to this
orbifold produces the flow
\begin{equation}
\BC^2/\BZ_{10(-3)}\rightarrow\BC^2/\BZ_{2(-1)}\oplus\BC^2/\BZ_{4(-3)}
\end{equation}
It is simple to see this from (\ref{data2}), just by splitting the data
into two parts, with the common vector being $(2,-1)$ and using an
appropriate $SL(2,Z)$ transformation. 
In this example, there is a second way to split the data. Note that the blowing up 
of the point of intersection of the $j$th and the $j+1$st $\BC\BP^1$ is described, in 
the toric language, by the following change in the corresponding continued 
fraction \cite{fulton}
\begin{equation}
[a_1,a_2,\cdots~,a_r]\rightarrow[a_1,a_2,\cdots~,(a_j+1),1,(a_j+1),\cdots~,a_r]
\label{split2}
\end{equation}
This corresponds to inserting a vector $v$ between $v_j$ and $v_{j+1}$ such 
that $v=v_j + v_{j+1}$. One way to do this is to modify the toric data to 
the following
\begin{equation}
{\cal T}~=~\pmatrix{1&0&-1&-2&-7& \cr 0&1&2&3&10}\equiv 
\pmatrix{1&0&-1&-3&-2&-7& \cr 0&1&2&5&3&10} 
\end{equation}
The self intersection numbers from the above data is seen to be
$[2,3,1,5]$ according to (\ref{split2}). Now, in an obvious way, we can
split the toric data into two parts, which can be recognised as corresponding 
to $\BC^2/\BZ_{5(3)}$ and $\BC^2/\BZ_{5(1)}$. 

In general, there will be more than one way to effect the above splitting, 
corresponding to the number of ways in which one can change the continued fraction,
as in (\ref{split2}). These are related to perturbations of the CFT by
products of generators of the chiral ring \cite{hkmm}. 

At this point, let us mention that the same analysis can be done for toric
singularities of the form $\BC^3/\Gamma$. Consider the singularities of the
form (\ref{threefold}), where for simplicity, we set $k=1$. We will refer to this
class of singularities as $\left[\frac{1}{n}\left(1,1,m\right)\right]$. These 
orbifolds are generally non-supersymmetric, and for these cases, localisation of the 
tachyons require that $m$ is even. A few of the non-supersymmetric cases will be 
treated in the next subsection. Here, we will briefly discuss the simplest supersymmetric
case, $\left[\frac{1}{3}\left(1,1,1\right)\right]$ \cite{asp},\cite{dgm}. 
(As is well known, when  $n$ is not a prime number, these orbifold will have non-isolated 
singularities. These, in particular, include the cases when $n$ is even. We will come back 
to this class of singularities in a while). Recall \cite{asp},\cite{oda} that for the
three-fold quotient singularity of the form $\left[\frac{1}{n}\left(1,k,m\right)\right]$, 
the toric fan is given by a cone in a three-dimensional lattice, 
generated by the following one dimensional cones
\begin{equation}
v_1=\left(ne^1-ke^2-me^3\right),~~~~~v_2=e^2,~~~~~v_3=e^3
\label{cones3}
\end{equation}
where $\left(e^1,e^2,e^3\right)$ are the three unit vectors that form a 
basis for the three dimensional lattice. The blowup of the singularity is 
effected by adding points in the toric diagram which are linear combinations 
of the above vectors, with weights determined by the action of the discrete
group (for details, the reader is referred to \cite{asp}). These internal vectors 
are of the form
\begin{equation}
u_j = \sum_{i=1}^3 t_iv_i,~~~~~~0\leq t_i\leq 1
\label{internalray}
\end{equation}
For our example $\left[\frac{1}{3}\left(1,1,1\right)\right]$, The one dimensional cones 
are given by
\begin{eqnarray}
v_1=\left(3,-1,-1\right),~~~v_2=\left(0,1,0\right),~~~v_3=\left(0,0,1\right)
\end{eqnarray}
In this case, we can add one internal ray as in (\ref{internalray}), with 
weights $t_i=\frac{1}{3},~i=1,2,3$, which is given by the vector 
$\left(1,0,0\right)$. This is a blowup of the singularity 
$\left[\frac{1}{3}\left(1,1,1\right)\right]$. 

It is simple to generalise the procedure of \cite{hkmm} in studying this class of 
examples. Namely, we can study marginal perturbations, which correspond to perturbations 
by the generators of the chiral ring in these examples, and, sending the coefficients of
these marginal modes to infinity, we flow to orbifolds of lower rank. We can see 
it directly from the toric data, using the $SL(3,Z)$ automorphism of the three 
dimensional lattice, much like the two-fold examples. In the particular case of 
the orbifold  $\left[\frac{1}{3}\left(1,1,1\right)\right]$, using the added vector 
$(1,0,0)$, it is easy to see, in analogy with the two-fold examples, that the toric data 
can be split up into that for three simpler spaces, all of which can be identified with 
the flat space $\BC^3$. Similar analyses can be carried out for higher rank 
supersymmetric orbifolds, and in a method similar to the two-fold examples, we have 
checked from the toric data for these orbifolds, that they flow to lower rank 
supersymmetric orbifolds.  
 
We now move on to the D-brane gauge theory descrpition of the singularities
that we have considered.

\subsection{D-brane Gauge Theory}

In the supersymmetric case, the toric data corresponding to a certain quotient 
singularity can be succinctly described in terms of the gauge theory living on the 
D-brane probing the singularity. The gauge theory is constructed by following the
well known prescription of Douglas and Moore \cite{dm}. Such a gauge theory, in general, 
is described by its matter content and its interactions. While the former is specified 
by the D-term equations, the latter is given by the superpotential, which lead
to the  F-term constraints. The data can be combined in a form prescribed in \cite{dgm}
in order to extract geometric information of the singularity that the 
D-brane probes. 
 
An analogous procedure can be followed for the non-supersymmetric orbifolds
that we have been considering, purely by considerations of the bosonic subsector
of the theory in the classical limit. The analogy is, in a sense, clear. 
When the tachyon expectation values are small compared to the string scale, 
i.e in the substringy regime, we can set up an analogue of the prescription of 
\cite{dgm}, by examining the classical moduli space of the scalar fields. 
Of course, when generic twisted sectors are turned on, the probe analysis will 
be less useful. But for the moment, let us assume that we are working in a regime
where fluctuations are very small, and one can use the classical picture.

Our analysis is similar to the supersymmetric case. Let us illustrate
this by an example which will also serve to set up the notations that we will use 
later. Consider the non-supersymmetric orbifold $\BC^2/\BZ_{5(3)}$.
This orbifold does not have any supersymmetric subsectors, and hence any 
deformation of the theory will be purely tachyonic. In this case, the low energy 
theory will be an orbifold of the D-brane gauge theory in flat space, obtained in the 
usual way \cite{dm} by the action of the discrete group on the coordinates and the 
Chan-Paton indices. By considering the classical scalar potential, one can write down 
the equivalents of the D and F terms. 

The projection of the fields is as in \cite{dm}, and the quiver diagram can be
obtained by standard techniques. This quiver diagram encodes information about the 
$U(1)^5$ charges of the unprojected fields. This can be written as a matrix $\Delta$, 
where one of the overall $U(1)$s denoting the centre of mass motion of the branes, 
is omitted,
\begin{equation}
\Delta~=~\pmatrix{ -1 & 0 & 0 & 0 & 1 & -1 & 0 & 1 & 0 & 0 \cr 1 & \
-1 & 0 & 0 & 0 & 0 & -1 & 0 & 1 & 0 \cr 0 & 1 & -1 & 0 & 0 & 0 & 0 & \
-1 & 0 & 1 \cr 0 & 0 & 1 & -1 & 0 & 1 & 0 & 0 & -1 & 0 \cr  } 
\end{equation}
The analogues of the F-terms are those necessary for the vanishing of the term 
$|[X^1,X^2]|^2$ at the minimum of the classical scalar potential. These terms, of the
form $[X^1,X^2]=0$ are not all independent. In fact, one can check that they
can be solved in terms of six independent fields, which we denote collectively
by $v_j, j=1\cdots 6$. The solution can be expressed in terms of a matrix $K$, 
such that the original fields of the theory, which we denote by $X_i$, can be 
expressed in terms of the six independent fields $v_j$ as $X_i~=~\prod_j v_j^{K_{ij}}$. 
In this example, the (transpose of) the matrix $K$ is given by 
\begin{equation} 
K^t~=~
\pmatrix{ 1 & 0 & 0 & 0 & 0 & 0 & -1 & \
-1 & 0 & 0 \cr 0 & 1 & 0 & 0 & 0 & 0 & 0 & -1 & \
-1 & 0 \cr 0 & 0 & 1 & 0 & 0 & 0 & 0 & 0 & -1 & \
-1 \cr 0 & 0 & 0 & 1 & 0 & 0 & 1 & 1 & 1 & 0 \cr 0 & 0 & 0 & 0 & 1 & 0 & 0 \
& 1 & 1 & 1 \cr 0 & 0 & 0 & 0 & 0 & 1 & 1 & 1 & 1 & 1 \cr  }
\label{cone}
\end{equation}
We now introduce a new set of fields, $p_{\alpha}, \alpha=1,\cdots c$, which are
the physical fields in the linear sigma model corresponding to this orbifold. The reason
for this transformation is to avoid the singularities that may arise, due to negative 
entries in (\ref{cone}). Given the matrix $K$, we calculate its dual cone $T$, which is 
composed of the set of vectors dual to $K$, i.e ${\vec K}.{\vec T}\geq 0$, following the 
algorithm given in \cite{fulton}. Then, the dual cone defines our new set of fields 
$p_{\alpha}$, where the relation between the independent variables $v_i$ of the original 
theory and these fields is given by $v_i~=~\prod_{\alpha}p_{\alpha}^{T_{i\alpha}}$.
In our case, the dual cone is given by
\begin{equation}
T~=~
\pmatrix{ 0 & 0 & 0 & 0 & 0 & 0 & 0 & 0 & 1 & 1 & 1 & 1 \cr 0 & 0 & 0 & 0 & 0 & 1 & \
1 & 1 & 0 & 0 & 0 & 1 \cr 0 & 0 & 0 & 1 & 1 & 0 & 0 & 0 & 0 & 0 & 1 & 1 \cr \
0 & 0 & 1 & 0 & 0 & 0 & 0 & 1 & 0 & 1 & 0 & 1 \cr 0 & 1 & 0 & 0 & 1 & 0 & \
1 & 0 & 0 & 0 & 0 & 1 \cr 1 & 0 & 0 & 1 & 0 & 1 & 0 & 0 & 1 & 0 & 1 & 0 \
\cr  }
\label{dualcone53}
\end{equation}
Since the number of new fields introduced is more than the number of independent
fields in the original theory, we need to introduce a new gauge 
group (corresponding to a number of $C^*$ actions) in order
to eliminate redunduncies. We can calculate the charges of the new fields
$p_{\alpha}$ under this new gauge group by using conditions of gauge invariance.  
Let us call the matrix denoting these charges as $V_1$. In addition, we can determine
the matrix of charges of these new fields under the original $U(1)^4$, which we
call $V_2$. This matrix $V_2$, which is the equivalent of the D-term in the gauge theory
with the new fields, will, contain the analogues of the FI parameters. Writing
the matrices without the coloumn of FI parameters, concatenating $V_1$ and $V_2$, and 
taking the kernel of the resulting matrix gives us the geometric data for the resolution 
of the singularity \cite{dgm}, which, in this case, is found be the matrix
\begin{equation}
{\cal T}~=~
\pmatrix{ 1 & 0 & -1 & -3 \cr 0 & 1 & 2 & 5 \cr  }
\end{equation} 
after the elimination of all the repeated coloumns. This is the expected result for the 
toric data of the orbifold $\BC^2/\BZ_{5(3)}$. \footnote{By constructing invariant 
variables in terms of the linear sigma model fields (as elaborated in the appendix for 
the orbifold $\BC^2/\BZ_{8(3)}$, the equation for this singularity is given by
$z^5=x^2y$ in appropriate gauge invariant variables $x,~y,~z$.}  

In this example, there are no supersymmetric subsectors, and the decay of this
space proceeds entirely by tachyonic deformations. In the next subsection, we will study
thse decays from the point of view of the inverse toric procedure \cite{fhh}, which is
essentially an algorithm to evaluate D-brane gauge theory configurations starting from
the toric data that the brane probes, i.e the opposite process of what we have
described above. For the moment, having set up the necessary conventions, let us study, 
in a similar fashion, the example of the orbifold $\BC^2/\BZ_{8(3)}$, in which we can use 
marginal deformations of the theory to split the space. We will find that the D-brane
toric data, in these cases, contains a novel element: it encodes the data for a marginal
deformation, in addition to the usual data for the singularity.

The analysis proceeds in complete analogy with the previous example, and we have 
relegated the details of the calculations in the appendix. The final result for the toric
geometry data for a D-brane probing the singularity $\BC^2/\BZ_{8(3)}$ are contained
in the coloumns of the matrix 
\begin{equation}
{\cal T}~=~
\pmatrix{ 1 & 0 & -1 & -1 & -3 \cr 0 & 1 & 4 & 3 & 8 \cr  } 
\label{eightthree}
\end{equation}
From the analysis of the previous section, we get the following
flow pattern
\begin{equation}
\BC^2/\BZ_{8(3)}\rightarrow\BC^2/\BZ_{4(1)}\oplus\BC^2/\BZ_{4(-3)}
\label{contra1}
\end{equation}
In agreement with \cite{aps}.
Notice that while the usual toric data for this example is given by the 
continuted fraction $[3,3]$ with the vectors of the toric fan being given by
$(1,0),(0,1),(-1,3),(-3,8)$, the toric data obtained from the D-brane probe 
gives the continued fraction $[4,1,4]$ with the added vector $(-1,4)$ which
corresponds to a marginal deformation by the fourth twisted sector, which, as
can be checked, is a supersymmetric subsector. Hence, we see that the toric 
data of the D-brane gauge theory contains an additional point which is blown up,
this point corresponding to a marginal deformation of the theory. Figure 
(\ref{fig2}) give the toric diagram for the orbifold $\BC^2/\BZ_{8(3)}$. 
As we can see, the point corresponding to the marginal deformation lies on the
line joining the initial and final vectors of the cone. 

Continuing along the same lines, let us now proceed to analyse nother example, 
the non-supersymmetric orbifold $\BC^2/\BZ_{10(3)}$. The explicit
matrices are not presented here due to space constraints. The final result
for the toric data, from considerations of the D-brane gauge theory is
\begin{equation}
{\cal T}~=~
\pmatrix{ 1 & 0 & -1 & -1 & -2 & -3 \cr 0 & 1 & 5 & 4 & 7 & 10 \cr  }
\label{contra2}
\end{equation} 
Whereas the original Hirzebruch-Jung singularity in this case is given by
$[4,2,2]$, we see that the toric data given by the D-brane gauge theory is 
$[5,1,3,2]$, with an additional point blown up, corresponding to a deformation by the 
fifth twisted sector, which is supersymmetric. Once again, we can obtain the split 
in the toric data in accordance with \cite{aps}. 

Let us now turn to the class of examples where the orbifolding group is of odd order.
One such example is the non-supersymmetric orbifold $\BC^2/\BZ_{9(5)}$ which has 
marginal deformations in the third twisted sector. The result for the toric data 
obtained from the D-brane gauge theory is, in this example,
\begin{equation}
{\cal T}~=~
\pmatrix{1&0&-1&-1&-5\cr 0&1&3&2&9\cr}
\label{ninefive}
\end{equation}
This corresponds to the continued fraction $\frac{9}{5}=[3,1,6]$ and, comparing with
the original continued fraction for this singularity, which is given by $[2,5]$, we see
that the D-brane toric data once again provides us with the correct marginal
deformation, corresponding to turning on the point $(3,-1)$. The explicit matrices for 
this example are presented in the appendix.

Let us mention at this point that in \cite{hkmm}, it was 
argued that decays of singularities of the form (\ref{contra}) may not
be allowed in the CFT description for large values of $l$. In the cases that
we have discussed in (\ref{contra1}) and (\ref{contra2}), $l$ is 
sufficiently small, and the results are consistent with the $g_{cl}$ conjecture 
of HKMM. However, purely from a D-brane probe analysis, there is no way to 
rule out the large $l$ flows.  We believe that a study of higher dimensional 
examples might shed more light into this apparent contradiction.

We are now ready to discuss some examples of non-supersymmetric 3-fold
orbifolds. The supersymmetric 3-folds were briefly discussed before, and as
we have mentioned, they can be treated in the same way as supersymmetric
2-folds, following HKMM. Let us now discuss the gauge theory data for 
D-branes probing 3-fold orbifolds. We will show that once again, the toric data 
encodes the information about the marginal deformations of the theory, as in the
two-fold examples. The low energy theory is again a quiver gauge theory, and one can 
proceed to determine the resolutions of these orbifolds along the lines of \cite{aps},
using the analogues of the FI parameters as before. The calculations are entirely similar 
to the two-fold cases, and we will not include the details. We present some 
results below. 

Consider first the non-supersymmetric orbifold given by
$\left[\frac{1}{5}(1,1,2)\right]$. The classical D-brane world volume theory is
a gauge theory $U(1)^5$ gauge theory in this case, and the toric data can
be calculated following the procedure of \cite{dm}, and is given by
\begin{equation}
{\cal T}~=~
\pmatrix{1&0&0&-1&-2\cr 0&1&0&0&-1\cr 0&0&1&3&5}
\label{5112}
\end{equation}
This is expected, as from (\ref{cones3}), we see that the toric
fan, in this example, is generated by the one dimensional cones given by the
vectors $(5,-1,-2),~(0,1,0),~(0,0,1)$. The vector $(1,0,0)$ corresponds to
adding an internal ray with the weight vector $\zeta = \left(\frac{1}{5},
\frac{1}{5},\frac{2}{5}\right)$, and the vector $(3,0,-1)$ is the internal
ray with the weight vector $\zeta^3$. 

Let us now go over to our next example, $\left[\frac{1}{6}(1,1,2)\right]$. 
There are non-isolated singularities in this case, since the rank of the orbifolding 
group is non-prime. We can, however, construct the D-brane gauge theory data as 
in the previous examples. This example is interesting, because  this orbifold has 
marginal deformations, corresponding to the third twisted sector, which corresponds
to $\left[\frac{1}{2}\left(1,1,2\right)\right]$ (noting that with fermions included, 
the integers in the brackets are defined modulo $4$, we see that this is a 
supersymmetric subsector). The final result for the toric data for this case is
\begin{equation}
{\cal T}~=~
\pmatrix{1&0&0&-1&-1&-2\cr 0&1&0&0&0&-1\cr 0&0&1&3&4&6}
\label{6112}
\end{equation}
Let us consider eq. (\ref{6112}) in some details. In this case, the toric fan
is generated by the vectors $(6,-1,-2),~(0,1,0),~(0,0,1)$. The weight
vector for the orbifold action, as follows from eq. (\ref{internalray}) 
is $\zeta = \left(\frac{1}{6}, \frac{1}{6},\frac{2}{6}\right)$, and the vector
$(1,0,0)$ corresponds to adding the internal ray with these weights. The vector 
$\left(4,0,-1\right)$ is the internal ray corresponding to $\zeta^4$. Similarly, 
the vector $\left(3,0,-1\right)$ corresponds to the action of $\zeta^3$ on the
vectors of the toric fan (note that the integers appearing in the weight vector is 
defined modulo $6$), and the latter, as we have pointed out, is a marginal deformation. 
Thus, as in the $\BC^2$ example, the D-brane gauge theory data corresponding to a 
non-supersymmetric three-fold orbifold contains an additional point, which, as
before, corresponds to a marginal deformation. 

As a final example, consider the orbifold $\left[\frac{1}{8}(1,1,2)\right]$, the fourth 
subsector of which has marginal deformations. The result for the toric data in this case is 
\begin{equation}
{\cal T}~=~
\pmatrix{1&0&0&-1&-1&-2\cr 0&1&0&0&0&-1\cr 0&0&1&4&5&8}
\label{8112}
\end{equation} 
Similar to our previous example, we see that the vector $\left(4,0,-1\right)$
corresponds to a deformation by the fourth subsector of the theory, which is a 
supersymmetric subsector. 

In all the above three-fold orbifold examples, the methods of the previous subsection 
can be used to split the toric data, and study the RG flows. These can also be understood
by generalising the procedure of \cite{hkmm}. These flows will involve perturbing by 
various relevant and marginal operators, as in the two-fold case. We will, however, 
leave a complete study of the same for a future publication. 

\subsection{The Inverse Toric Procedure}

Now that we have discussed examples of D-brane world-volume gauge theories for 
branes probing non-supersymmetric orbifolds from the toric geometry point of view, 
we can ask the question of the existance of toric duality \cite{fhh} for non-supersymmetric 
orbifolds. Indeed, the end point of closed string tachyon condensation in the class 
of examples that we have studied is always a supersymmetric orbifold, but 
these orbifolds have opposite supersymmetry \cite{aps} compared to the usual supersymmetric
orbifolds. Hence, it is possible that we might learn something new by considering toric 
duals of the gauge theories that we have been considering so far. 
 
It is well known that an interesting aspect of world volume gauge theories of 
D-branes probing toric singularities is the so called toric duality, first proposed in 
\cite{fhh}.  Simply stated, this duality principle states that there can be more than one 
D-brane gauge theory that flows in the IR to the same universality class i.e share the same 
toric description. This, in particular, followed from the inverse toric algorithm developed 
in \cite{fhh} which allows one to read off the world volume gauge theory data of D-branes 
from the toric data of the singularity that it probes. It is a useful tool in the context 
of non-supersymmetric orbifolds, as we will show now. Let us start by briefly reviewing 
the inverse toric algorithm of \cite{fhh}. 
 
The computation of the geometric data of the singularity probed by a D-brane
was developed in \cite{dgm}. As we have already pointed out in the beginning of the
last subsection, this method consists of concatenating the D and F-terms
of the (supersymmetric) gauge theory of the D-brane world volume into a matrix that
describes the dual cone of the toric variety that the brane probes. 
In our discussion in the previous subsections, we have carried out this procedure 
for non-supersymmetric orbifolds also, and we have shown that interestingly, the result 
often describes a non-minimal resolution of the orbifold, with additional points in 
the toric diagram corresponding to marginal perturbations of the theory. 

The inverse toric algorithm, on the other hand, involves embedding the original
orbifold singularity into one of higher rank, and then determining the partial
resolutions of the latter in order to reach the lower rank orbifold in question. It
turns out that in the process, one might discover new gauge theories that are 
torically dual to the original one.

The above procedure is similar to the one that we followed in determining the flows of
non-supersymmetric D-brane orbifold gauge theories to supersymmetric ones (or for 
that matter from supersymmetric gauge theories for orbifolds of higher ranks to 
those of lower rank). The important point to note here is that, in the language of
\cite{fhh}, the final supersymmetric orbifold theories that are obtained from the 
flows can be embedded in {\it non-supersymmetric} orbifolds. Then, giving vev's 
to some of the original fields of the theory can be equivalently stated in 
terms of the fields of the linear sigma model corresponding to the orbifold. Knowing 
exactly which fields are to be resolved in order to flow from a non-supersymmetric 
to a supersymmetric orbifold, we can determine the conditions on the    
classical moduli space (corresponding to the analogues of the FI parameters) 
of the original theory. To set the notation, let us illustrate this with an example.

We choose the simple model of $\BC^2/\BZ_{8(3)}$. From the brane probe 
point of view, we have seen that the flows seen by the D-brane are those
for which one can construct marginal deformations of the gauge theory. 
This would, in particular, correspond to choosing a subset of the analogues
of the Fayet-Illiapoulos parameters and setting linear combinations of 
them to zero. Any such combination, provided the relevant deformation is 
marginal, would result in an orbifold of lower rank, starting from the parent orbifold. 
In the inverse toric procedure, this would imply that we resolve a certain number
of points in the toric diagram in a way consistent with turning on of
these parameters in the gauge theory. For this particular example,
the number of physical fields was found to be $24$ (see eqn. (\ref{dualcone83})
of the appendix), and gauge invariant polynomials are constructed in terms 
of these. 

\begin{figure}
\centering
\epsfxsize=3.5in
\hspace*{0in}\vspace*{.2in}
\epsffile{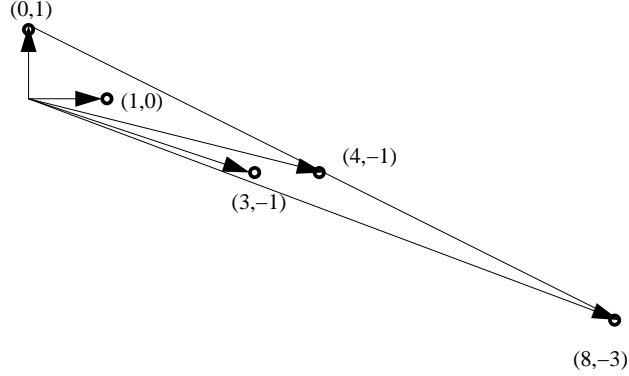}
\caption{\small The toric diagram for the singularity $\BC^2/\BZ_{8(3)}$, 
corresponding to eq. (\ref{contra1}). The toric diagram for $\BC^2/\BZ_{4(1)}$,
which consists of the vectors $(4,-1),~(1,0),~(0,1)$ can be embedded into this.} 
\label{fig2}
\end{figure}

The equation for the singularity can be constructed in terms of the gauge
invariant parameters of the original theory. In this example, there are three 
gauge invariant combinations that we can form (the others being equivalent to these), 
and the expressions for these are presented in the appendix. As expected, when expressed
in terms of the linear sigma model fields, they satisfy a relation of the form $xy=z^4$. 
We note that from the toric data of eq. (\ref{contra1}), the toric diagram for the 
singularity $\BC^2/\BZ_{4(1)}$ can be embedded into that of $\BC^2/\BZ_{8(3)}$, 
as shown in fig. (\ref{fig2}) and using this, we can determine the fields in the linear 
sigma model that need to be resolved in order to effect the flow 
$\BC^2/\BZ_{8(3)} \to \BC^2/\BZ_{4(1)}$ using the techniques of \cite{fhh}.

Equivalently, from the analysis of section 2 (following \cite{aps}), it is clear 
that in order to retain a $\BZ_4$ symmetry at the end, we need to give vevs to four of 
the fields which are then resolved, and the remaining four correspond to the residual 
$\BZ_4$ symmetry. Let us choose these to be the fields $X_{12},X_{34},X_{56},X_{78}$. 
From the expression for these fields in terms of the linear sigma model fields 
(\ref{dualcone83}), it can be seen that in terms of these, this resolution implies that 
the fields that we need to retain in this process are $p_1,p_2,p_4,p_6, p_{11},p_{15}$, 
and the rest are eliminated. \footnote{This was one of the problems addressed in \cite{fhh}. 
In general, the resolution of a node of a toric diagram implies the resolution of more 
than one field in the linear sigma model. In the method of \cite{fhh}, the fields that have 
to be resolved are determined by the embedding of the toric diagram, in this case, 
they are equivalently determined by looking at the vevs given to the fields of the 
original theory.} From this data, we can construct the reduced charge matrix of the fields 
by removing the relevant coloumns from the dualcone of eq. (\ref{dualcone83}), and
by suitably rearranging its rows so as to make sure that this reduced dualcone is
appropriate for the reduced toric data, obtained after removing the points $(8,-3)$
and $(3,-1)$ from the original toric data. If we also add the coloumn containing the 
analogues of the FI parameters in the original charge matrix $V_2$ 
(in the notation immediately following eq. (\ref{dualcone53})), the reduced 
charge matrix includes these, and the final result for this matrix is 
\begin{equation}
Q_{total}=\pmatrix{ 1 & -1 & -1 & -1 & -1 & 1 & 0 \cr 0 & 0 & -1 & 1 & 0 & 0 
& \lambda_5 + \lambda_6 \cr 
0 & 0 & 0 & -1 & 1 & 0 & \lambda_3 + \lambda_4 
\cr 0 & 1 & 0 & 0 & -1 & 0 & \lambda_1 + \lambda_2 \cr} 
\label{charge41}
\end{equation}
Where the $\lambda$s are the analogues of the FI parameters. Let us mention that 
from this matrix, we can immediately read off the region of parameter space that we 
are dealing with. From eq. (\ref{charge41}), it is clear that in order to effect the 
turning on of the vevs of the original fields in the theory in the manner that we have 
done in the previous paragraph, we need to set one of the seven FI parameters to a  
finite positive value (and hence remove it from the toric data), and choose the 
linear combinations of the other six to be zero. This choice of the classical parameter 
space can also be obtained directly by following the procedure of \cite{fhh}, i.e by 
performing a Gaussian row reduction on the original charge matrix, and tuning the fields 
so that one gives non-zero vevs to some fields while staying in the physical region of 
the parameter space where the FI parameters $\lambda_i\geq 0$. From (\ref{charge41}), 
after following the procedure of \cite{fhh}, the charge matrix (corresponding to the quiver 
diagram) of the resulting gauge theory can be found to be 
\begin{equation}
d~=~\pmatrix{0&1&-1&0&0&1&-1&0\cr
1&-1&0&0&1&-1&0&0\cr
-1&0&0&1&-1&0&0&1\cr
0&0&1&-1&0&0&1&-1\cr}
\label{delta41}
\end{equation}
which is the charge matrix for the orbifold $\BC^2/\BZ_{4(1)}$ where the last
row represents the trivial $U(1)$ corresponding to the centre of mass motion of the brane.

In deriving the above result, we have assumed that the first row in 
(\ref{charge41}) corresponds to the F-term and the other three rows correspond
to the D-terms of the final gauge theory. However, as has been pointed out in
\cite{fhh}, the division of the matrix (\ref{charge41}) into D and F terms is 
actually arbitrary. Writing the matrix (\ref{charge41}) without specifying the FI 
parameters, we can make a different choice for the D and the F-terms. Let us study
this in some more details. If, in (\ref{charge41}), we chose the first two
rows as specifying the F-terms, and the other two as corresponding to the D-terms,
a calculation in the lines of \cite{fhh} shows that we get a charge matrix
\begin{equation}
d=\pmatrix{0&-1&1&0&1&-1\cr
1&0&-1&1&-1&0\cr
-1&1&0&-1&0&1\cr}
\label{dual}
\end{equation}
where, as before, we have added the trivial $U(1)$. The dual matrix $K$ of the kernel 
of the F-term matrix (which, in this case, is the first two rows of the matrix in 
(\ref{charge41})) is given by 
\begin{equation}
K=\pmatrix{0&0&0&1&1&2\cr
0&0&1&0&1&0\cr
0&1&0&0&0&1\cr
1&0&0&1&0&0\cr}
\label{superpot83}
\end{equation}
Since the nullspace of $K$ has dimension $2$, we expect two relationships 
between the coloumns of $K$ for this theory. They are given by 
$\left(X_1X_3X_6=X_2X_4X_5\right)$, and $\left(X_1X_5=X_3X_4\right)$. 
Now, from the charge matrix in 
eq. (\ref{dual}), we can construct the gauge invariant quantities, $a=X_1X_2X_3$, 
$b=X_4X_5X_6$, $c=X_2X_3X_4$, $d=X_1X_5X_6$, $e=X_1X_2X_5$, $f=X_3X_4X_6$, $g=X_1X_3X_6$ 
and $h=X_2X_4X_5$. Using the first of the relations between the coloumns, we  
obtain $ab=g^2$, which is the equation for the singularity $\BC^2/\BZ_{4(1)}$.
\footnote{Note that the orbifold $\BC^2/\BZ_{4(1)}$ has the opposite supersymmetry
as compared to the usual orbifold $\BC^2/\BZ_{4(-1)}$. While the equation for the 
latter singularity is $xy=z^4$, the former is given by $xy=z^2$, in some appropriate
gauge invariant variables.} The same relation can be derived by considering other
algebraic relations between the gauge invariant variables. For eg., it can be seen
by using both the relations between the coloumns of $K$ in (\ref{superpot83}) that
another possible algebraic relation between the gauge invariant variables is $de=g^2$. 
Integrating the relations between the coloumns of $K$ 
to form the superpotential will in general need the introduction of new (presumably
chargeless) fields in the theory, and in general non zero values of these new
fields might lead us into different branches of the moduli space. We leave a detailed
discussion of this issue for the future.  

While the above example seems to suggest some sort of toric duality 
\cite{toricduality}, with the matrix (\ref{dual}) 
corresponding to a $U(1)^3$ D-brane world volume gauge theory, which is 
different from the $U(1)^4$ theory of (\ref{delta41}), let us point out a few 
caveats. In the non-supersymmetric case that we have been considering, our 
discussion is restricted to vanishing string coupling, and therefore, 
we cannot make a statement about the two gauge theories of (\ref{delta41})
and (\ref{dual}) as being dual (that flow to a common
fixed point in the IR) at this point. The example given above should be thought
of as relating two gauge theories that have the same classical moduli space. 
Nevertheless, we believe that it might be possible to make a stronger assertion about 
the quantum corrections and the IR behaviour of the two theories in (\ref{delta41}) 
and (\ref{dual}), and work is in progress in this direction.   

Further, both the theories in eqs. (\ref{delta41}) and (\ref{dual}) have chiral
fermions and hence will have anomalies at the quantum level. Anomaly cancellation
in the usual supersymmetric case can be checked following the procedure advocated in 
\cite{dm}. We have not performed such explicit checks here. 
 
At this stage, let us point out that it is important in these examples that the toric 
data for the final (supersymmetric) singularity that we want to reach can be embedded 
in the original (non-supersymmetric) one. This can be done when the original orbifold 
has marginal deformations. For the $\BC^2$ orbifolds, these deformations correspond, in 
the toric diagram, to additional points on the edges of the toric diagram, as in
fig. (\ref{fig2}). This process of embedding is however, not possible for 
non-supersymmetric orbifolds without marginal deformations, and we do not expect the 
inverse toric procedure to be of much use in those cases. In this paper, we have studied 
the inverse toric algorithm for two-fold orbifolds only. It will be very interesting to 
carry out this analysis for the case of three-fold orbifolds. 
 
\subsection{Weighted Projective Spaces}

Clearly, we expect the above analysis to go over to the case of 
higher dimensional orbifolds, of the form $\BC^n/\Gamma$. Already, for
$n=3$, we expect a much richer structure than two-fold orbifolds. 
One possible route to investigate would be the behaviour of brane probes
on weighted projective spaces, considered in \cite{vafa}. Consider, for
example, the supersymmetric orbifold $\BC^3/\BZ_4$ with the action of the
discrete group on the coordinates being given by
\begin{equation}
\left(Z^1,~Z^2,~Z^3\right) \rightarrow \left(\omega Z^1,~\omega Z^2,~\omega^2 Z^3\right)
\end{equation}
Where $\omega$ is a fourth root of unity. 
The blowup of the origin of this orbifold will correspond to the weighted
projective space $\BC\BP^2_{1,1,2}$. A similar analysis can be done for the 
non-supersymmetric orbifolds in $\BC^3$. One can ask if D-brane probes
will be useful in studying decays of the blowups of such orbifolds. Vafa \cite{vafa}
has considered the mirror Landau-Ginzburg models corresponding to such 
weighted projective spaces. Clearly, the analogues of these are expected
to be seen in the corresponding Gepner models, and it might be possible to study 
these decays from the geometric point of view as presented here. This will involve 
the construction of the analogue of a Poincare polynomial in lines with the 
supersymmetric case. We expect the toric geometry of brane probes 
to be an useful tool of analysis in such cases.

\section{Summary}

In this paper, we have carried out an investigation of the condensation of 
closed string tachyons in non-supersymmetric orbifold theories, in the 
sub-stringy regime. As tools, we have used the D-brane probe method developed in 
\cite{aps} and toric geometry methods, which are intimately related to the chiral ring 
techniques of \cite{hkmm}. We have provided a partial classification of flow patterns
in orbifolds of the form $\BC^2/\BZ_n$ using quiver techniques. We have seen
from a consideration of the D-brane gauge theory that it probes the correct 
singularity for the non-supersymmetric orbifolds, and, where appropriate, the toric 
data arising from the D-brane gauge theory provides additional points corresponding to 
marginal deformations.

We have also examined a few examples of three-fold orbifolds, and found that
even in these cases, the toric data of the D-brane probing these singularities 
encodes the information about the marginal deformations by adding additional points
to the toric diagram.

Further, we have applied the inverse toric algorithm to a class of non-supersymmetric
orbifolds, which have marginal deformations in some of the twisted sectors. We have
shown that this algorithm can be applied to these non-supersymmetric orbifolds, and we have
initiated a discussion of toric duality in the same. Whereas in \cite{toricduality},
dual D-brane gauge theories were studied for supersymmetric orbifolds, an extension of
our results in this paper might give examples of such duality in orbifolds which have the 
opposite supersymmetry compared to the usual supersymmetric ones.  
 
It would be interesting to further this investigation in a number of directions. 
Even though the probe method has limited applicability, it might 
prove to be useful for analysis of higher dimensional non-supersymmetric orbifolds, 
In particular, the inverse toric method can be used to ask questions about
a more general underlying structure of toric duality. Further, as we have 
mentioned, a direct application of our methods can be made in the study of
the decay of weighted projective spaces. Consider, for example, the model
$\BP^4_{1,1,2,2,2}$. Methods of \cite{roanyau} and \cite{vafalg}
can be used to deduce the singularity structure for this manifold in a 
Landau-Ginzburg framework, and one can calculate the various forms 
that need to be blown up in order to resolve the singularities of this space. It would
be interesting to consider this example in the light of its realisation
as the blowup of a corresponding (non-supersymmetric) orbifold of the form 
$\BC^5/\Gamma$.  

Also, it would be interesting to extend this type of analysis in the
case of orbifolds of product groups, for eg. of the form $\BC^3/\BZ_m\times
\BZ_n$. It follows from \cite{wittenphases} that for such theories, the 
phase diagram is more complicated than those for $\BZ_n$ orbifolds. 
It would be very interesting to study closed string tachyon condensation
in these theories, by toric geometry methods. We leave these issues for a future 
publication. 

\section{Acknowledgements}
We would like to thank S. Mukhopadhyay and K. Ray for useful discussions and
for collaboration during the initial part of this project. We would also like
to thank K. S. Narain, G. Thompson, M. S. Narasimhan, D. Goswami and T. E. 
Venkata Balaji for helpful discussions.

\section{Appendix}
\subsection{Toric data for the singularity $\BC^2/\BZ_{8(3)}$}
The Matrix $K^t$ is given by
\begin{equation}
K^t=
\pmatrix{ 1 & 0 & 0 & 0 & 0 & 0 & 0 & 0 & 0 & -1 & -1 & -1 & -1 & \
-1 & 0 & 0 \cr 0 & 1 & 0 & 0 & 0 & 0 & 0 & 0 & 0 & 0 & -1 & -1 & -1 & -1 & \
-1 & 0 \cr 0 & 0 & 1 & 0 & 0 & 0 & 0 & 0 & 0 & 0 & 0 & -1 & -1 & -1 & -1 & \
-1 \cr 0 & 0 & 0 & 1 & 0 & 0 & 0 & 0 & 0 & 1 & 1 & 1 & 0 & 0 & 0 & 0 \cr 0 \
& 0 & 0 & 0 & 1 & 0 & 0 & 0 & 0 & 0 & 1 & 1 & 1 & 0 & 0 & 0 \cr 0 & 0 & 0 & \
0 & 0 & 1 & 0 & 0 & 0 & 0 & 0 & 1 & 1 & 1 & 0 & 0 \cr 0 & 0 & 0 & 0 & 0 & \
0 & 1 & 0 & 0 & 0 & 0 & 0 & 1 & 1 & 1 & 0 \cr 0 & 0 & 0 & 0 & 0 & 0 & 0 & \
1 & 0 & 0 & 0 & 0 & 0 & 1 & 1 & 1 \cr 0 & 0 & 0 & 0 & 0 & 0 & 0 & 0 & 1 & \
1 & 1 & 1 & 1 & 1 & 1 & 1 \cr  } 
\end{equation}
The dual cone is generated by
\begin{equation}
\pmatrix
{ 0 & 0 & 0 & 0 & 0 & 0 & 0 & 0 & 0 & 0 & 0 & 0 & 0 & 0 & 0 & 0 & 1 & 1 & \
1 & 1 & 1 & 1 & 1 & 1 \cr 0 & 0 & 0 & 0 & 0 & 0 & 0 & 0 & 0 & 0 & 1 & 1 & \
1 & 1 & 1 & 1 & 0 & 0 & 0 & 0 & 0 & 0 & 1 & 1 \cr 0 & 0 & 0 & 0 & 0 & 0 & \
1 & 1 & 1 & 1 & 0 & 0 & 0 & 0 & 0 & 1 & 0 & 0 & 0 & 0 & 1 & 1 & 0 & 1 \cr \
0 & 0 & 0 & 0 & 0 & 1 & 0 & 0 & 0 & 1 & 0 & 0 & 0 & 1 & 1 & 0 & 0 & 1 & 1 & \
1 & 0 & 0 & 0 & 1 \cr 0 & 0 & 0 & 0 & 1 & 0 & 0 & 0 & 1 & 0 & 0 & 1 & 1 & \
0 & 0 & 0 & 0 & 0 & 0 & 1 & 0 & 1 & 1 & 1 \cr 0 & 0 & 0 & 1 & 0 & 0 & 0 & \
1 & 0 & 0 & 0 & 0 & 0 & 0 & 1 & 1 & 0 & 0 & 1 & 0 & 1 & 0 & 1 & 1 \cr 0 & \
0 & 1 & 0 & 0 & 0 & 0 & 0 & 0 & 1 & 0 & 0 & 1 & 1 & 0 & 1 & 0 & 1 & 0 & 0 & \
0 & 1 & 0 & 1 \cr 0 & 1 & 0 & 0 & 0 & 0 & 0 & 1 & 1 & 1 & 0 & 1 & 0 & 0 & \
1 & 0 & 0 & 0 & 0 & 1 & 0 & 0 & 0 & 1 \cr 1 & 0 & 0 & 0 & 0 & 0 & 1 & 0 & \
0 & 0 & 1 & 0 & 0 & 0 & 0 & 1 & 1 & 0 & 0 & 0 & 1 & 1 & 1 & 0 \cr  } 
\label{dualcone83}
\end{equation}
The charge matrix under the original $U(1)^7$ is
\begin{equation}
Q=
\pmatrix{ -1 & 0 & 0 & 0 & 0 & 0 & 0 & 1 & \
-1 & 0 & 0 & 0 & 0 & 1 & 0 & 0 \cr 1 & -1 & 0 & 0 & 0 & 0 & 0 & 0 & 0 & \
-1 & 0 & 0 & 0 & 0 & 1 & 0 \cr 0 & 1 & -1 & 0 & 0 & 0 & 0 & 0 & 0 & 0 & \
-1 & 0 & 0 & 0 & 0 & 1 \cr 0 & 0 & 1 & -1 & 0 & 0 & 0 & 0 & 1 & 0 & 0 & \
-1 & 0 & 0 & 0 & 0 \cr 0 & 0 & 0 & 1 & -1 & 0 & 0 & 0 & 0 & 1 & 0 & 0 & \
-1 & 0 & 0 & 0 \cr 0 & 0 & 0 & 0 & 1 & -1 & 0 & 0 & 0 & 0 & 1 & 0 & 0 & \
-1 & 0 & 0 \cr 0 & 0 & 0 & 0 & 0 & 1 & -1 & 0 & 0 & 0 & 0 & 1 & 0 & 0 & \
-1 & 0 \cr  } 
\end{equation}
From these, the toric data given in (\ref{eightthree}) can be calculated.
The gauge invariant combinations in terms of the physical fields (corresponding
to the coloumns of the dual cone in (\ref{dualcone83}) are given by
\begin{eqnarray}
X&=&X_{12}X_{23}X_{34}X_{45}X_{56}X_{67}X_{78}X_{81}\nonumber\\
&=&p_2\cdots p_7p_8^3p_9^3p_{10}^4p_{11}p_{12}^3p_{13}^3p_{14}^3p_{15}^4p_{16}^4
p_{17}p_{18}^3p_{19}^3p_{20}^4p_{21}^3p_{22}^4p_{23}^4p_{24}^8\nonumber\\
Y&=&Y_{14}Y_{47}Y_{72}Y_{25}Y_{58}Y_{83}Y_{36}Y_{61}\nonumber\\
&=&p_1^8\left(p_2\cdots p_7\right)^3p_8p_9p_{10}^4p_{11}^3p_{12}p_{13}p_{14}
p_{15}^4p_{16}^4 p_{17}^3p_{18}p_{19}p_{20}^4p_{21}p_{22}^4p_{23}^4\nonumber\\
Z&=&X_{12}Y_{25}X_{56}Y_{61}\nonumber\\
&=&p_1^2p_2\cdots p_9p_{10}^2p_{11}\cdots p_{14}p_{15}^2p_{16}^2p_{17}p_{18}p_{19}
p_{20}^2p_{21}p_{22}^2p_{23}^2p_{24}^2
\end{eqnarray}
  
\subsection{Toric data for the singularity $\BC^2/\BZ_{9(5)}$}
For this example, the matrix $K^t$ is given by
\begin{equation}
\pmatrix{ 1 & 0 & 0 & 0 & 0 & 0 & 0 & 0 & 0 & 0 & -1 & -1 & -1 & \
-1 & 0 & 0 & 0 & 0 \cr 0 & 1 & 0 & 0 & 0 & 0 & 0 & 0 & 0 & 0 & 0 & -1 & \
-1 & -1 & \
-1 & 0 & 0 & 0 \cr 0 & 0 & 1 & 0 & 0 & 0 & 0 & 0 & 0 & 0 & 0 & 0 & -1 & \
-1 & -1 & \
-1 & 0 & 0 \cr 0 & 0 & 0 & 1 & 0 & 0 & 0 & 0 & 0 & 0 & 0 & 0 & 0 & -1 & \
-1 & -1 & \
-1 & 0 \cr 0 & 0 & 0 & 0 & 1 & 0 & 0 & 0 & 0 & 0 & 0 & 0 & 0 & 0 & -1 & \
-1 & -1 & \
-1 \cr 0 & 0 & 0 & 0 & 0 & 1 & 0 & 0 & 0 & 0 & 1 & 1 & 1 & 1 & 1 & 0 & 0 & \
0 \cr 0 & 0 & 0 & 0 & 0 & 0 & 1 & 0 & 0 & 0 & 0 & 1 & 1 & 1 & 1 & 1 & 0 & \
0 \cr 0 & 0 & 0 & 0 & 0 & 0 & 0 & 1 & 0 & 0 & 0 & 0 & 1 & 1 & 1 & 1 & 1 & \
0 \cr 0 & 0 & 0 & 0 & 0 & 0 & 0 & 0 & 1 & 0 & 0 & 0 & 0 & 1 & 1 & 1 & 1 & \
1 \cr 0 & 0 & 0 & 0 & 0 & 0 & 0 & 0 & 0 & 1 & 1 & 1 & 1 & 1 & 1 & 1 & 1 & \
1 \cr  } 
\end{equation}
The dual cone, $T$ in this example, is generated by
\begin{equation}
\pmatrix{
   0 & 0 & 0 & 0 & 0 & 0 & 0 & 0 & 0 & 0 & 0 & 0 & 0 & 0 & 0 & 0 & 0 & 0 & 1 & 1 & 1 & 1 & 1 \cr 0 & 
   0 & 0 & 0 & 0 & 0 & 0 & 0 & 0 & 0 & 0 & 0 & 0 & 0 & 1 & 1 & 1 & 1 & 0 & 0 & 0 & 0 & 1 \cr 0 & 0 & 
   0 & 0 & 0 & 0 & 0 & 0 & 0 & 0 & 1 & 1 & 1 & 1 & 0 & 0 & 0 & 0 & 0 & 0 & 0 & 0 & 1 \cr 0 & 0 & 0 & 
   0 & 0 & 0 & 0 & 1 & 1 & 1 & 0 & 0 & 0 & 0 & 0 & 0 & 0 & 0 & 0 & 0 & 0 & 1 & 1 \cr 0 & 0 & 0 & 0 & 
   0 & 1 & 1 & 0 & 0 & 0 & 0 & 0 & 0 & 0 & 0 & 0 & 0 & 1 & 0 & 0 & 1 & 0 & 1 \cr 0 & 0 & 0 & 0 & 1 & 
   0 & 0 & 0 & 0 & 0 & 0 & 0 & 0 & 1 & 0 & 0 & 1 & 0 & 0 & 1 & 0 & 0 & 1 \cr 0 & 0 & 0 & 1 & 0 & 0 & 
   0 & 0 & 0 & 0 & 0 & 0 & 1 & 0 & 0 & 1 & 0 & 0 & 0 & 0 & 0 & 1 & 1 \cr 0 & 0 & 1 & 0 & 0 & 0 & 0 & 
   0 & 0 & 1 & 0 & 1 & 0 & 0 & 0 & 0 & 0 & 1 & 0 & 0 & 0 & 0 & 1 \cr 0 & 1 & 0 & 0 & 0 & 0 & 1 & 0 & 
   1 & 0 & 0 & 0 & 0 & 1 & 0 & 0 & 0 & 0 & 0 & 0 & 0 & 0 & 1 \cr 1 & 0 & 0 & 0 & 0 & 1 & 0 & 1 & 0 & 
   0 & 1 & 0 & 0 & 0 & 1 & 0 & 0 & 1 & 1 & 0 & 1 & 1 & 0 \cr  }
\end{equation}
Finally, the charge matrix of the fields under the original $U(1)^8$ is
\begin{equation}
\pmatrix{ -1 & 0 & 0 & 0 & 0 & 0 & 0 & 0 & 1 & \
-1 & 0 & 0 & 0 & 1 & 0 & 0 & 0 & 0 \cr 1 & \
-1 & 0 & 0 & 0 & 0 & 0 & 0 & 0 & 0 & \
-1 & 0 & 0 & 0 & 1 & 0 & 0 & 0 \cr 0 & 1 & \
-1 & 0 & 0 & 0 & 0 & 0 & 0 & 0 & 0 & \
-1 & 0 & 0 & 0 & 1 & 0 & 0 \cr 0 & 0 & 1 & \
-1 & 0 & 0 & 0 & 0 & 0 & 0 & 0 & 0 & \
-1 & 0 & 0 & 0 & 1 & 0 \cr 0 & 0 & 0 & 1 & \
-1 & 0 & 0 & 0 & 0 & 0 & 0 & 0 & 0 & \
-1 & 0 & 0 & 0 & 1 \cr 0 & 0 & 0 & 0 & 1 & \
-1 & 0 & 0 & 0 & 1 & 0 & 0 & 0 & 0 & \
-1 & 0 & 0 & 0 \cr 0 & 0 & 0 & 0 & 0 & 1 & \
-1 & 0 & 0 & 0 & 1 & 0 & 0 & 0 & 0 & \
-1 & 0 & 0 \cr 0 & 0 & 0 & 0 & 0 & 0 & 1 & \
-1 & 0 & 0 & 0 & 1 & 0 & 0 & 0 & 0 & -1 & 0 \cr  } 
\end{equation}
From this data, one obtains (\ref{ninefive}).

\end{document}